\documentclass[conference]{IEEEtran}
\IEEEoverridecommandlockouts
\usepackage{cite}
\usepackage{amsmath,amssymb,amsfonts}
\usepackage{graphicx}
\usepackage{textcomp}
\usepackage{xcolor}
\usepackage[hidelinks]{hyperref}

\usepackage{ragged2e}

\usepackage{listings}

\usepackage{algorithm}
\usepackage{algpseudocode}
\usepackage{amsmath}

\def\BibTeX{{\rm B\kern-.05em{\sc i\kern-.025em b}\kern-.08em
    T\kern-.1667em\lower.7ex\hbox{E}\kern-.125emX}}
\begin{document}

\title{Privacy-Preserving and Trustworthy Localization in an IoT Environment
}

\author{
\IEEEauthorblockN{1\textsuperscript{st} Guglielmo Zocca\IEEEauthorrefmark{1}\IEEEauthorrefmark{2}, 2\textsuperscript{nd} Omar Hasan\IEEEauthorrefmark{2}}

\IEEEauthorblockA{\IEEEauthorrefmark{1}Department of Information Engineering and Computer Science, University of Trento, Italy}

\IEEEauthorblockA{\IEEEauthorrefmark{2}Institut National des Sciences Appliquées (INSA) de Lyon, University of Lyon, France}

\IEEEauthorblockA{guglielmo.zocca@studenti.unitn.it, omar.hasan@insa-lyon.fr}
}

\maketitle

\begin{abstract}
The Internet of Things (IoT) is increasingly prevalent in various applications, such as healthcare and logistics. One significant service of IoT technologies that is essential for these applications is localization. The goal of this service is to determine the precise position of a specific target. The localization data often needs to be private, accessible only to specific entities, and must maintain authenticity and integrity to ensure trustworthiness.
IoT technology has evolved significantly, with Ultra-Wide Band (UWB) technology enhancing localization speed and precision. However, IoT device security remains a concern, as devices can be compromised or act maliciously. Furthermore, localization data is typically stored centrally, which can also be a point of vulnerability.
Our approach leverages the features of a permissioned blockchain, specifically Hyperledger Fabric, to address these challenges. Hyperledger Fabric's collection feature ensures data privacy, and its smart contracts (chaincode) enhance trustworthiness. We tested our solution using a network of devices known as CLOVES, demonstrating robust performance characteristics with UWB technology. Additionally, we evaluated our approach through an indoor localization use case.
\end{abstract}


\section{Introduction}
\label{introduction}

IoT technology is increasingly becoming an integral part of our daily lives and business operations. It is embedded in home appliances as network chips, found in security cameras, and widely utilized in the logistics field to track transported products. In localization use cases, IoT devices can be used to determine the position of individuals indoors for various analytical purposes. For instance, in a museum, visitors can be given a device that communicates with anchor devices to track their path throughout the exhibits, providing valuable insights into visitor behavior and preferences.

Despite the incredible range of applications, IoT devices often rely on a central server, which may become a single point of security failure. Blockchain technology can enhance the architecture by providing integrity, auditability, and authenticity. However, public blockchains allow all data to be accessible to everyone, which is not suitable for applications requiring privacy and access control. For instance, in certain localization applications, it is necessary to ensure that sensitive data, such as the position of a target, is only accessible to specific clients and not to all users and peers connected to the blockchain.

For this project, we employ Hyperledger Fabric, a private, permissioned blockchain, to achieve the required levels of performance and security. This blockchain framework ensures data privacy through its collection feature, which supports data segmentation and controlled access. Consequently, experimental data can be limited to specific participants, safeguarding confidentiality. Hyperledger Fabric's collections allow a selected group of entities within a channel to endorse, commit, or query private data. Each collection includes two primary elements: the private data, transmitted peer-to-peer solely to authorized entities, and the hash of the private data, which is endorsed, ordered, and recorded in the ledger's world state on every peer within the channel.

These elements guarantee that data remains private while still being endorsed and validated by all blockchain nodes. Furthermore, access policies can be linked to collections, permitting only designated client groups to access the collection data.

Additionally, Hyperledger Fabric supports the creation of smart contracts, known as chaincode, using general-purpose programming languages, which enhances flexibility and streamlines the development of business logic. These smart contracts run on the blockchain, providing assurance of trust and integrity in their execution.
In this paper, we utilize smart contracts for the trust management of IoT devices, building on the methodology defined by Dedeoglu et al. \cite{b1}.

This combination of features makes Hyperledger Fabric an ideal choice for securely managing and processing IoT data while maintaining privacy and performance. 

Our contributions to improving IoT application privacy are as follows: 1. We leverage the concept of Hyperledger collections to ensure that IoT device data is accessible only to clients participating in a specific experiment. By defining policies for these collections, we can enforce stringent access controls, ensuring data privacy and confidentiality within the blockchain network; 2. We implement access control checks within the chaincode to further privatize data within the same experiment. These checks differentiate between admin and user clients, ensuring that sensitive data is accessible only to those with the appropriate permissions.

All of this technology is tested and evaluated using both general and ad-hoc testing frameworks. For instance, Hyperledger Caliper is employed to assess the performance of chaincode execution. Additionally, we conducted IoT network performance tests using the CLOVES \cite{b2} (Communication and Localization Testbed for Validation of Embedded Systems) testbed at the University of Trento. This testbed is accessible to anyone who subscribes to the service, providing a practical and cost-effective solution for evaluating business logic in a network of devices.

The results of these tests confirm that our work's primary objectives are achieved: ensuring the privacy of device data among different groups of clients and types of clients, and the correct management of IoT device trust. However, the performance results indicate that the current implementation is not yet optimal for localization processes. The chaincode and consensus processing require significant time, highlighting the need for increased parallelization and other techniques to improve efficiency.

This project resulted in the creation of a prototype \cite{b3} — a step towards developing a secure architecture for processing position data from devices in a secure and private manner. The use case involves a client who can register with the system and conduct localization experiments alongside other clients. The client must build his/her network of devices and application business logic upon the security, functionality, and privacy provided by the blockchain.

The paper is organized as follows: Section \ref{our-goals}: Description of our goals; Section \ref{building-blocks}: Description of the building blocks of our system;  Section \ref{our-proposal}: Explanation of the different aspects of our system; Section \ref{operations}: Explanation in more detail of the security and localization algorithms developed; Section \ref{use-case}: An example of deployment and use of our system. Section \ref{evaluation}: Design and description of the tests executed; Section \ref{results}: Discussion of the tests results; Section \ref{related-work}: Discussion of related work; Section \ref{conclusion}: Conclusion of the paper.

\section{Our Goals}
\label{our-goals}

The goals of our work are as follows:
\begin{enumerate}
    \item \textbf{Create a prototype system for IoT localization}: This setup will include anchor devices to determine the position of a target entity.
    \item \textbf{Implement trust management for IoT devices}: Specifically, this system will assess the reputation of each device, evaluate the confidence of its observations, and consider observations from other devices to determine which data should be used for accurately calculating the target's location.
    \item \textbf{Ensure a trustworthy, secure, and decentralized system}: Design a system that remains reliable and secure even in the presence of malicious devices or blockchain nodes. This will be achieved through the use of blockchain technology and a robust trust management system for IoT devices.
    \item \textbf{Preserve user privacy within the system}: Implement privacy measures to ensure that only participants of an experiment can access its data. Additionally, further privatize data within the same experiment by distinguishing between admin and user clients, ensuring sensitive data is accessible only to those with appropriate permissions.
    \item \textbf{Test in an IoT envirmonent}: Test the localization operations of devices within a real IoT network to achieve more accurate results.
\end{enumerate}

Our project takes inspiration from the paper ``A Trust Architecture for Blockchain in IoT'' \cite{b1} by Dedeoglu et al. We build on their architecture, composed of gateways and devices. Additionally, we adopted the concept of utilizing a permissioned blockchain for enhanced performance and implemented a trust system for devices to address potential threats posed by malicious entities. However, our work improves user privacy and adds access control to the data.

%The aim of our research is to enhance a blockchain structure by incorporating data privatization features, drawing inspiration from the paper ``A Trust Architecture for Blockchain in IoT'' \cite{b1}. We aim to demonstrate these novel characteristics through an example application—a prototype illustrating an end-to-end process from devices to users for calculating the position of a target.

%Our goal is to achieve data privatization within the architecture, ensuring that experiment application data stored in the blockchain are accessible only to those responsible for the experiment or authorized individuals. We strive to achieve this property without compromising the inherent advantages of blockchain technology, including auditability, integrity, and authenticity.

\section{Building Blocks}
\label{building-blocks}

\subsection{Hyperledger Fabric}
 
Hyperledger Fabric is an open-source, enterprise-grade permissioned distributed ledger technology (DLT) platform, specifically designed for enterprise use. It offers significant advantages over other blockchain platforms with its highly modular and configurable architecture, enabling innovation, versatility, and optimization across various industries such as banking, finance, insurance, healthcare, human resources, supply chain, and digital music delivery. Information about Hyperledger Fabric can be found on its official website \cite{b9}\cite{b11}. The definitions and details reported in this section are drawn from the above sources.

\subsubsection{Main Components}

Hyperledger Fabric consists of several modular components, including:

\begin{itemize}
    \item \textbf{Peer nodes}: Maintain the ledger and propose transactions.
    \item \textbf{Ordering service node}: Establishes consensus on the order of transactions and broadcasts blocks to peers. This service is logically decoupled from the peers that execute transactions and maintain the ledger.
    \item \textbf{Membership Service Provider (MSP)}: Associates network entities with cryptographic identities, linking entities to either user clients or admin clients.
    \item \textbf{Peer-to-peer gossip service}: Disseminates blocks produced by the ordering service to other peers.
    \item \textbf{Smart contracts (chaincode)}: Run within a container environment (e.g., Docker) for isolation.
    \item \textbf{Endorsement and validation policy enforcement}: Configurable independently for each application.
    \item \textbf{Certificate Authority (CA)}: Creates cryptographic materials, such as digital certificates and private keys, enabling secure identification and communication among network entities.
\end{itemize}

\subsubsection{Smart Contracts}

Hyperledger Fabric supports smart contracts, known as ``chaincode'', which can be written in general-purpose programming languages such as Java, Go, and Node.js. This flexibility allows enterprises to develop smart contracts using their existing skills without the need for additional training in domain-specific languages (DSL). Smart contracts function as trusted distributed applications, gaining security and trust from the blockchain and peer consensus.

\subsubsection{Permissioned Characteristics}

Fabric operates as a permissioned network, where participants are known to each other rather than being anonymous. Although participants may not fully trust each other (e.g., competitors in the same industry), a governance model can be built on existing trust. This reduces the risk of malicious code being introduced through smart contracts. All actions, such as submitting transactions, modifying network configurations, or deploying smart contracts, are recorded on the blockchain in accordance with an established endorsement policy.

\subsubsection{Consensus Protocol}

One of Fabric's key differentiators is its support for pluggable consensus protocols, allowing customization to fit specific use cases and trust models. Fabric can use consensus protocols that do not require a native cryptocurrency to incentivize mining or fuel smart contract execution. This reduces significant risk vectors and operational costs, as the platform can be deployed with the same cost structure as a regular distributed system.

\subsubsection{Privacy}

Hyperledger Fabric ensures confidentiality through its channel architecture and private data feature. The MSP allows for the definition of organizations, which are logical sets of peers and users. To maintain privacy for data transactions and smart contracts from other organizations on a channel, a new channel can be created for the relevant organizations. However, managing separate channels introduces administrative overhead.

To address this, Fabric offers private data collections, enabling a defined subset of organizations within a channel to endorse, commit, or query private data without creating a separate channel. This simplifies administration and supports scenarios requiring data privacy within a shared channel. A collection includes:

\begin{itemize}
    \item \textbf{Private data}: Transmitted peer-to-peer exclusively to authorized organizations, stored in a private state database on authorized peers, and accessible through chaincode. The ordering service does not access the private data.
    \item \textbf{Hash of private data}: Endorsed, ordered, and written to the ledger's world state on every peer within the channel. This hash serves as proof of the transaction for state validation and audit purposes.
\end{itemize}

These elements ensure that data remains private while being endorsed and validated by all endorsement and order nodes.

\subsubsection{Our Reasoning for Selecting Hyperledger Fabric as a Building Block} Hyperledger Fabric enhances privacy through its unique properties. In contrast to systems where all experiment data submitted to the blockchain is universally accessible to all clients, as discussed in the paper by Dedeoglu et al. \cite{b1}, Hyperledger Fabric enables the segregation of data based on experiments and client types. This segregation is achieved through the definition of collections and enforcement by smart contracts (chaincode). Each collection can specify experiment data and access policies, restricting access exclusively to designated clients. The chaincode verifies whether a client is an administrator or user within their organization, thereby managing their access privileges within the same experiment. Overall, Hyperledger Fabric ensures robust data separation and privacy control.

\section{Our Proposal}
\label{our-proposal}

The structure of the system is composed of five elements.

\subsection{IoT Network}
This component represents the network of IoT devices responsible for capturing external information. The testbed utilized for testing the IoT network is CLOVES \cite{b2} (Communication and Localization Testbed for Validation of Embedded Systems). CLOVES is a large-scale public testbed dedicated to ultra-wideband (UWB) \cite{b12} technology. This testbed is located within the Department of Information Engineering and Computer Science at the University of Trento. The project utilizes the evb1000 device type.

In our project, these devices perform a ranging process with a specific target. Some devices within the CLOVES network serve as anchors, while others act as the target for localization purposes. The concept revolves around anchors conducting two-way ranging with the target, with each distance calculation being logged by the device. A parsing program then extracts crucial data from the log, including the device ID, target ID, calculated distance in millimeters, and the confidence level assigned to the distance.
The confidence level is determined based on the Received Signal Strength Indicator (RSSI) from the target. If the signal power from the target exceeds a certain threshold, the device assigns maximum confidence. Conversely, if the signal power falls below another threshold, the device assigns minimum confidence. For cases falling outside of these thresholds, confidence is calculated using a formula based on the received RSSI.
This approach is designed to account for variations in signal strength, where weaker signals may indicate greater distance or obstacles between the target and device, potentially leading to less precise measurements. This methodology draws inspiration from Dedeoglu et al \cite{b1}.

After parsing, the data is encrypted using XOR encryption \cite{b10}. The encryption key is a symmetric key, stored both in the device and within the corresponding device entity in the blockchain. Before encryption, the data is hashed, and the resulting digest is sent along with the encrypted data. This approach ensures confidentiality through encryption and integrity through hashing. XOR encryption was chosen for its simplicity and performance.
To ensure authenticity, the gateway application checks that the ID of the received data matches the ID of the device entity in the blockchain. This method was also chosen for its simplicity and performance.

In the specific case of the prototype, the three phases—ranging, parsing, and encryption—are separate, sequential processes connected through files. Each phase uses the file generated from the previous phase. Future projects building on this research may aim to integrate these phases more closely to create a more realistic and efficient prototype.

\subsection{Blockchain} 
This component is implemented using Hyperledger Fabric, where it stores all data related to devices and the target. The data is categorized into collections associated with specific organizations. Put simply, each administrator can create their own organization within Fabric and associate it with collections containing the target and devices. Consequently, only members of a given organization can access the associated data. These organizations and collections establish a form of access control and data privatization.
Additionally, the blockchain hosts the chaincode necessary for various functionalities within the system. This includes calculations for target position by the gateway application, access to the target by users, and administrative tasks related to maintaining the data and system structure.
In the prototype, the blockchain component is simulated using Docker containers.

\subsubsection{Test Network (Prototype Network)} Hyperledger provides developers with a test network that includes all the essential elements for Hyperledger Fabric, along with some example applications. In this research, this network is customized to meet the specific requirements of the project. From a physical perspective, the network comprises the following components:
\begin{itemize}
\item 4 order nodes that create blocks using the PBFT (Practical Byzantine Fault Tolerance) consensus algorithm, which provides countermeasures against potential malicious entities within the network.
\item 1 peer node that handles the endorsement, proposal of transactions, and maintenance of collections for Organization 1.
\item 1 peer node that handles the endorsement, proposal of transactions, and maintenance of collections for Organization 2.
\end{itemize}
In the prototype, instead of using a service for the creation of cryptographic material, a command named ``cryptogen'' is utilized. These components are simulated using Docker containers.
The logical structure of the test network consists of the following components:
\begin{itemize}
\item 2 organizations. Each organization is a logical unit representing an experiment, comprising its own devices, target, administrators, users, peer nodes, and order nodes.
%\begin{figure}[t]%
%\FIG{\includegraphics[width=0.9\textwidth]{img/Logical_test.png}}
%{\caption{Logical point of view of the prototype blockchain}
%\label{fig5}}
%\end{figure}
\item 1 admin client credential for every organization.
\item 1 user client credential for every organization.
\item 1 world state of the blockchain for all organizations.
\item 2 collections for every organization. The collections of an organization can only be accessed by clients or peers that belong to that specific organization, ensuring data privacy and access control within the blockchain network.

The devices collection contains device data and can only be accessed by Admin entities (client, application, or peer) within the same organization. A device entity in the collection includes the following fields: ID of the device; key for decrypting data from the device; X coordinate of the device (relative to the network map); Y coordinate of the device (relative to the network map); distance observed by the device; Confidence level of the device in its observation; evidence value, representing the support for the device’s observation from neighboring devices; reputation of the device; trust level in the device's observation; IDs of neighboring devices. 

The target collection contains target data and can be accessed by User and Admin entities (client, application, or peer) within the same organization. However, only Admin entities can modify the data. A target entity in the collection includes the following fields: ID of the target; X coordinate of the target (relative to the network map); Y coordinate of the target (relative to the network map); timestamp of the target's position update; status indicator showing whether the target's position is updated or not.
\end{itemize}
\subsubsection{Chaincode}
The business logic of the blockchain is implemented in Golang due to its efficiency, usability, and simplicity. Within this chaincode, a single smart contract named ``PositionContract'' is defined and used by all organizations. Some data passed to the contract methods are kept in a transient state to ensure these data are not visible in the blockchain transaction corresponding to that contract call. This transient option ensures confidentiality of the transaction data. Overall, this contract provides the following functionalities:
\begin{itemize}
    \item Create a device in the devices collection by an admin of the same organization of collection.
    \item Update device neighborhood or decryption key by an admin of the same organization of device.
    \item Create the target in the target collection by an admin of the same organization of collection.
    \item Update device observation and confidence by an admin of the same organization of device.
    \item Update device reputation and device observation evidence and trust by an admin of the same organization of device.
    \item Delete a device from the devices collection by an admin of the same organization of device.
    \item Delete the target from the target collection by an admin of the same organization of the target.
    \item Calculate the target position by an admin of the same organization of the target.
    \item Read the target by an admin or a user of the same organization of the device.
    \item Read all ids of the device in the devices collection by an admin of the same organization of collection.
    \item Read a device by an admin of the same organization of the device.
\end{itemize}

\subsection{Gateway Node and App} 
This component represents the nodes to which the devices connect to transmit data. These nodes host an application in Golang with administrative permissions. This application handles all communication operations with the blockchain and its chaincode. Its responsibilities include updating and creating device entities, as well as calculating and storing the position of the target in the blockchain.
In the prototype, the gateway application simulates to receive data from the devices, but only time considerations are taken into account.
\subsection{User Client and App} It is an entity with a user account within a specific organization. It has limited access, primarily to the target's position via a User Golang application. Access to this account is granted by an organization administrator.
\subsection{Admin Client and App} It an entity that possesses an administrative account within a designated organization. Through an Admin Golang application, it has the authority to modify device properties stored in the blockchain (decryption key or neighborhood), access device and target data, and create new device entities. The individual holding this account is often the same person who establishes the network of devices and initiates the gateway application to conduct experiments within the secure environment.

\section{Operations}
\label{operations}

In this section, specific operations executed within the system are elaborated in detail. The algorithms presented in subsections D, E, F, and G and their descriptions are inspired by the paper by Dedeoglu et al \cite{b1}.
\subsection{Encryption of an observation from a device} 
This algorithm involves applying the XOR operation \cite{b10} between a single character key and every character of the string passed to the algorithm to encrypt the string. The XOR operator's property allows for the same set of instructions to be used for decryption as well, making the algorithm simple and fast.

\begin{algorithm}[H]
\caption{Encrypt or Decrypt using XOR}
\begin{algorithmic}[1]
\Function{encryptDecrypt}{$\text{inpString}[ ]$, $\text{key}$}
    \State $\text{xorKey} \gets \text{key}$ 
    \State $\text{len} \gets \text{length of } \text{inpString}$
    \For{$i \gets 0$ \textbf{to} $\text{len} - 1$}
        \If{$i \neq \text{len} - 1$}
            \State $\text{inpString}[i] \gets \text{inpString}[i] \oplus \text{xorKey}$ 
        \EndIf
    \EndFor
\EndFunction
\end{algorithmic}
\end{algorithm}
\subsection{Hashing of the observation from device} 
The hash function is applied to the observation from the device before encryption, and the resulting digest is sent along with the encrypted data to the appropriate gateway.
\subsection{Two-way ranging}
The two-way ranging process involves calculating the period T1, which is the time between when the device sends the first message and when it receives the response from the target. Additionally, the period T2 is calculated, which is the time between when the target receives the message and when it sends the response to the device. Then, the time of flight (ToF) between the device and target is determined using the formula $ToF = (T1 - T2)/2$. Finally, the distance is calculated as $distance = ToF * speed of light$.
\subsection{Calculation of the confidence} 
The confidence level is determined based on the Received Signal Strength Indicator (RSSI) received from the target device. If the signal power from the target exceeds a certain threshold value RSSIUP, the device assigns a maximum confidence level of 1. Conversely, if the signal power falls below another threshold value RSSILOW, the device assigns a minimum confidence level of 0.4. For cases falling between these thresholds, the confidence level is calculated using the formula $9/4 + rssval/40$, where rssval represents the RSSI value calculated. These choices are based on the understanding that weaker signals may indicate greater distance or obstacles between the target and device, leading to potentially less precise measurements.
\begin{algorithm}
\caption{Calculate the confidence of device observation}
\begin{algorithmic}[1]
    \State $\text{conf} \gets 0$
    \If{$\text{rssval} > \text{RSSIUP}$}
        \State $\text{conf} \gets 1$
    \Else
        \If{$\text{rssval} < \text{RSSIINF}$}
            \State $\text{conf} \gets 0.4$
        \Else
            \State $\text{conf} \gets \frac{9}{4} + \frac{\text{rssval}}{40}$
        \EndIf
    \EndIf
\end{algorithmic}
\end{algorithm}
\lstset{
    language=Go,
    basicstyle=\ttfamily\small,
    keywordstyle=\color{blue}\bfseries,
    commentstyle=\color{gray}\itshape,
    stringstyle=\color{red},
    numbers=left,
    numberstyle=\tiny\color{gray},
    frame=single,
    rulecolor=\color{black},
    xleftmargin=2em,
    xrightmargin=2em,
    aboveskip=1em,
    belowskip=1em,
    backgroundcolor=\color{lightgray!20},
    tabsize=4,
    breaklines=true,
    showstringspaces=false,
    captionpos=b
}
\subsection{Calculation of the evidence of a device observation} 
 The calculation leverages the correlation among device observations to compute the evidence component for observation trust. The evidence (Evi) for an observation is determined based on data received from neighboring devices, whose information is stored in the device entity on the blockchain. If a neighbor's observation supports a device's observation, it increases the device's evidence by an amount proportional to the neighbor's observation confidence. Conversely, if a neighbor's observation contradicts the device's observation, it decreases the device's evidence by a value proportional to the neighbor's observation confidence. The final device confidence is adjusted according to the number of neighbors (num).

In this project, a neighbor's observation regarding the distance to the target supports the device's observation if the triangle formed by the edge between the two devices and the two edges between the devices and the target satisfies the triangle inequality. If this property is not upheld, it indicates that the distance calculated by these devices cannot pertain to the same target.
\begin{algorithm}
\caption{Calculate the evidence of a device obseervation}
\begin{algorithmic}[1]
    \State \textbf{Input:} Device data, Collection
    \State \textbf{Output:} Evidence of device $Evi$
    \State $Evi \gets 0$ \Comment{Evidence of device}
    \State $num \gets 0$ \Comment{Number of neighbors}
    \State $tmp \gets 0$
    \ForAll{$idN$ in $deviceData.Neigh$}
        \State $num \gets num + 1$
        \State $DeviceAsBytes, err \gets GetPrivateData(collection, idN)$
        \If{$err \neq nil$}
            \State \Return{Error: failed to get device: $err$}
        \ElsIf{$DeviceAsBytes == nil$}
            \State \Return{Error: this device does not exist: $idN$}
        \EndIf
        \State $deviceNeigh \gets \text{Unmarshal}(DeviceAsBytes)$
        \If{$err \neq nil$}
            \State \Return{Error: failed to unmarshal JSON: $err$}
        \EndIf
        \If{$NOT Respected Triangle Inequality$}
            \State $tmp \gets tmp + deviceNeigh.Conf$
        \Else
            \State $tmp \gets tmp - deviceNeigh.Conf$
        \EndIf
    \EndFor
    \State $Evi \gets tmp \times \left(\frac{1}{num}\right)$
\end{algorithmic}
\end{algorithm}
\subsection{Calculation of device reputation}
The calculation elucidates the intricate relationship between the trust level in an observation and the long-term reputation of its data source. A higher reputation of a node engenders greater trust in the node's observations. The reputation of a device evolves over time, with the guiding principle of reputation updates (based on the observation confidence and the evidence of neighbours observations)  being as follows: the magnitude of reputation reward or penalty should be proportional to the reported confidence.

When a device exhibits high confidence in its observation (i.e., Conf $\geq$ ThresholdConf) and the observation is corroborated by other nodes (i.e., Evi $\geq$ ThresholdEv), the device merits a substantial increase (PRH) in its reputation. Conversely, if a device provides observations with high confidence that are contradicted by other nodes, its reputation should undergo a significant decline. Similarly, rewards and penalties (PRL) for observations with low confidence should be less pronounced, i.e., PRL $<$ PRH.

In the context of this project, measures are implemented to prevent reputations from becoming excessively high or low. Specifically, reputation cannot be reduced below 0, nor can it exceed a maximum value (MaxRep).

\begin{algorithm}
\caption{Calculate the reputation of a device}
\begin{algorithmic}[1]
    \State \textbf{Input:} deviceData, deviceInput, Evi
    \State \textbf{Output:} Repu
    
    \State Repu $\gets$ deviceData.\text{Rep}

    \If {deviceData.\text{Conf} $\geq$ deviceInput.\text{ThreashConf} \textbf{and} Evi $\geq$ deviceInput.\text{ThreashEv}}
        \State Repu $\gets$ Repu + deviceInput.\text{PRH}
        \If {Repu $>$ deviceInput.\text{MaxRep}}
            \State Repu $\gets$ deviceInput.\text{MaxRep}
        \EndIf
    \EndIf

    \If {deviceData.\text{Conf} $<$ deviceInput.\text{ThreashConf} \textbf{and} Evi $\geq$ deviceInput.\text{ThreashEv}}
        \State Repu $\gets$ Repu + deviceInput.\text{PRL}
        \If {Repu $>$ deviceInput.\text{MaxRep}}
            \State Repu $\gets$ deviceInput.\text{MaxRep}
        \EndIf
    \EndIf

    \If {deviceData.\text{Conf} $\geq$ deviceInput.\text{ThreashConf} \textbf{and} Evi $<$ deviceInput.\text{ThreashEv}}
        \State Repu $\gets$ Repu - (deviceInput.\text{PRH} + 1)
        \If {Repu $<$ deviceInput.\text{MaxRep}}
            \State Repu $\gets$ deviceInput.\text{MaxRep}
        \EndIf
    \EndIf

    \If {deviceData.\text{Conf} $<$ deviceInput.\text{ThreashConf} \textbf{and} Evi $<$ deviceInput.\text{ThreashEv}}
        \State Repu $\gets$ Repu - (deviceInput.\text{PRL} + 1)
        \If {Repu $<$ deviceInput.\text{MaxRep}}
            \State Repu $\gets$ deviceInput.\text{MaxRep}
        \EndIf
    \EndIf
\end{algorithmic}
\end{algorithm}

\subsection{Calculation of the device observation trust}  Due to the interrelation between a device's observation confidence and evidence, and device reputation, the trustworthiness of a device's observation can be calculated as follows: $Trust = Confidence * Reputation * Evidence$.
\subsection{Calculation of target position} 
To determine the target's position, a form of multilateration is employed, involving the intersection of circumferences. In this project, the positions of the three most trusted devices are considered as the centers of these circumferences, with the calculated distances serving as their radii.

Initially, the intersection between the circumferences generated by the first two devices is computed. If no intersection occurs, the target's position cannot be calculated. If the intersection yields a single point, it becomes the target's position. In cases where two points are obtained, both are plugged into the analytical equation of the circumference defined by the third device ($x^2 + y^2 - r^2 = 0$). Subsequently, the point that yields a solution closest to 0 and is within the maximum error threshold (Thresh) is selected as the target's position. If the value derived from the expression $x^2 + y^2 - r^2$ exceeds the maximum error, the position is not calculated due to the anticipated high margin of error.

\section{Use Case}
\label{use-case}

In this section, n use case of deployment, implementation, and execution of the system is presented.
\subsection{Initialization} 
Initially, the Hyperledger Fabric infrastructure is deployed with several nodes and a channel, along with the contract ``PositionContract''. Next, a client registers as the admin of their organization on the blockchain, obtaining an admin account. Within the organization, the client establishes the necessary collections and specifies the peers that will be connected to the blockchain, either by creating new ones or utilizing existing ones.

The client proceeds to deploy the devices and target (The devices network), creating an application based on the prototype gateway application. This application includes the initial devices and the target. The admin then determines which peer or peers will serve as the gateway(s) and connects them to the devices.

Finally, the admin initiates the application on the gateway using own account and starts the firmware for ranging deployed in the devices. All operations performed on the gateway reference the contract already present in the blockchain.
\subsection{Processing} The devices perform two-way ranging with the target and send their results encrypted to the gateway. The gateway begins by checking the devices in the system. For each device, it receives and decrypts the data, buffering enough observations to compute distance and confidence averages, which are then updated on the blockchain.
Subsequently, the gateway updates the devices' reputation, as well as the evidence and trust for device observations. Finally, it computes the target's position and updates it on the blockchain. Once completed, the gateway returns to checking the devices in the system and capturing observations in the buffer, thus restarting the cycle.
\subsection{Admin Client} An admin can create and operate their own application based on the prototype admin application. This application establishes a connection with the ``PositionContract'' on the blockchain with the admin account. Through this interface, the admin gains control over certain aspects of the blockchain infrastructure. Specifically, the admin can read device data, access target data, update device information, and create new devices as needed.
\subsection{User Client} 
An admin has the authority to register another client with a user account. Subsequently, a user can independently create and operate their own application based on the prototype user application. This application establishes a connection with the ``PositionContract'' on the blockchain with the user account. Through this interface, the user gains the ability to access and retrieve the position of the target whenever it is updated.

\section{Evaluation}
\label{evaluation}

\subsection{Objectives}
The prototype undergoes analysis across three primary dimensions: correctness, privacy, and performance.
\begin{itemize}
\item \textbf{Correctness}: 
This topic focuses on rigorous testing to ensure the accuracy of calculations for device observation trust, target positioning, as well as the operations for reading and updating device data and target data.
\item \textbf{Privacy}: 
In this segment, the prototype is subjected to various scenarios to evaluate its response when privacy needs to be preserved.
\item \textbf{Performance}: 
In this section, the performance of the prototype is evaluated across different components, including the chaincode, applications, and the device network.
\end{itemize}

\subsection{Design}
For each topic discussed below, we will first outline the input parameters for the tests and then describe the tools and structure of the tests.

\subsubsection{Test Inputs}
The inputs are:
\begin{itemize}
\item \textbf{Correctness and privacy tests inputs}: For these tests, we use several test devices with IDs 1, 2, 3, 4, and 5, along with a test target with ID 7. The first three devices are designed to provide accurate observations for localizing a specific target with ID 7. See Table \ref{tab1} for the input data of the devices.
\begin{table}[htbp]
\caption{Application parameters for the tests}
\begin{center}
\begin{tabular}{|c|c|c|c|c|c|c|c|}
\hline
\textbf{Device} & \textbf{X} & \textbf{Y} & \textbf{Neigh} & \textbf{Decrkey} & \textbf{Conf} & \textbf{Dist} & \textbf{Rep} \\
\hline
1 & 3 & 2 & (2,3) & P & 1 & 4242 mm & 5 \\
2 & 10 & 4 & (1,3) & P & 1 & 4123 mm & 5 \\
3 & 5 & 8 & (2,1) &  P & 1 & 3162 mm & 5 \\
4 & 1 & 1 & (1,1) &  P & 8 & 4242 mm & 5 \\
5 & 3 & 2 & (3,2) &  P & 1 & 2 mm & 5 \\
\hline
\end{tabular}
\label{tab1}
\end{center}
\end{table}
The device 5 has the same position as device 1 to study the case of device 1 with incorrect observation. The device 4, instead, has a wrong confidence value. All these tests are in the context of organization 1. The values for the application parameters can be seen in table \ref{tab2}.

\begin{table}[htbp]
\caption{Application parameters for the tests}
\begin{center}
\begin{tabular}{|c|c|}
\hline
\textbf{Parameter}& \textbf{Value} \\
\hline
$Max_{\mathit{Conf}}$ & 1 \\
$Min_{\mathit{Conf}}$ & 0.4 \\
PRH & 2 \\
PRL & 1 \\
$Thresh_{\mathit{Conf}}$ & 0.7 \\
$Thresh_{\mathit{Ev}}$ & 0 \\ 
$Collection_{\mathit{Devices}}$ & DeviceAdmin1PrivateCollection \\
$Collection_{\mathit{Target}}$ & TargetOrg1PrivateCollection \\ $MaxError_{\mathit{Pos}}$ & 0.01 \\
$Max_{\mathit{Rep}}$ & 20 \\
$DimensionBatch_{\mathit{Obs}}$  & 6 \\
TimeReadObs & 100 ms \\
\hline
\end{tabular}
\label{tab2}
\end{center}
\end{table}

Those parameters that are also found in the paper of Dedeoglu et al. \cite{b1} have been assigned the same values as those presented in the paper for the sake of maintaining consistency with the experimental setup described in the paper.
\item \textbf{Performance tests inputs}: For these tests, all the configuration files (check config/core.yaml and Network/bft-confing/confitgx.yaml \cite{b3}) of the test network are considered. 
The specific devices and target data used as input are not crucial for evaluating these tests.
\end{itemize}
\subsubsection{Test Tools and Structure}
The tools and structure of the tests are:
\begin{itemize}
\item \textbf{Correctness tests structure and tools}: In this topic, the correctness of various application functionalities is tested using the Golang testing package. Some functionalities, such as device observation, evidence and trust updating, and device reputation updating, are tested in their test versions without considering performance strategies such as multithreading, as they are not pertinent to this test topic. The test functions can be found in the files ``Position$\_$test.go'' and ``Conf$\_$test.go''. Below is an outline of the tests:
\begin{itemize}
\item Test of accurate blockchain initialization.
\item Test of accurate addition of the devices observations and observation confidences.
\item Test of adding observations and observation confidences with a device with a wrong confidence value. It cannot be possible to insert the observation of the faulty device.
\item Test of accurate update of the observations evidence and trust, and update the of the devices reputations. Any device entity must have a reputation of 7 and evidence of 1.
\item Test of accurate calculation of the target position. 
\item Test of accurate retrieval of an updated target data as a user.
\item Test of retrieving the data of a target that has not been updated as a user. A user cannot read a target if it hasn’t been updated.
\item Test of updating the evidence and trust of observations, as well as the reputations of the devices. It includes a scenario where one device provides an incorrect distance measurement. The evidence and reputation of the faulty device are adjusted to -1 and 2.
\item Test of updating the position of the target. It includes a scenario where one device provides an incorrect distance measurement.
The test must indicate that the position cannot be calculated.
\item Test of updating the position of the target with 4 devices. One of the devices provides an incorrect distance measurement. The target position will be updated because the devices with the highest trust are considered.
\item Test of accurate update of the device decryption key or neighborhood by an Admin.
\item Test of accurate retrieval of the device data as an Admin.
\end{itemize}

\item \textbf{Privacy tests structure and tools}: In this topic, we test how the system handles potential privacy breaches across different application functionalities using the Go testing package. The tests are located in the files ``Position$\_$test.go'' and ``Conf$\_$test.go''. Below is an outline of the tests:
\begin{itemize}
\item Assess whether a user from organization 1 can access the devices collection of the same organization during the initialization process as a user of organization 1. The test should confirm that a user cannot access the devices collection.
\item Verify whether a client from organization 2 can access the devices collection of organization 1 through blockchain initialization as an admin of organization 2. The test should indicate that a client cannot access the devices collection of an organization to which it does not belong.
\item Examine whether a client from organization 2 can access the target collection of organization 1 by retrieving target data of organization 1 as an admin of organization 2. The test should indicate that a client cannot access the target collection of an organization to which it does not belong.
\end{itemize}
\item \textbf{Performance tests structure and tools}: The performance tests are executed in various environments and with diverse tools:
\begin{itemize}
\item \textbf{Application position computation}: 
In this test, the time taken to complete a cycle of target position computation by the gateway application is measured. This is done during a simulation of the gateway application, which also includes the addition of observations and confidence, and the updating of evidence, reputation, and trust.
\item \textbf{Ranging computation}: 
In this test, the average time taken to perform a ranging operation in the testbed environment between a device and the target is calculated. This calculation is derived from logs of the testbed experiment using a Python parsing code.
\item \textbf{Chaincode computations}: 
In the analysis of the chaincode operations performance using the Hyperledger Caliper tool, the throughput and the average time to execute the chaincode are computed for the following operations:
\begin{itemize}
\item Adding a device observation and device observation confidence.
\item Updating the evidence and trust of an observation, as well as the reputation of a device.
\item Position computation.
\item Reading the target.
\item Reading a device.
\item Updating a device by an Admin.
\end{itemize}
\end{itemize}
\end{itemize}
\subsubsection{Execution}
The procedure for executing the tests or simulation can be found on the GitHub page \cite{b3}. The GitHub repository provides a detailed explanation of the directory structure and contents of the prototype as well.

\section{Results}
\label{results}

In this section, the results of the tests are presented and discussed.

\subsection{Outcome}
\subsubsection{Correctness, Security and Privacy}
All these test types yielded positive results (PASS), verifying the system's intended correctness, security, and privacy.
\subsubsection{Performance}
See the results in Table \ref{tab3}. 
\begin{table}[htbp]
\caption{Performance tests results}
\begin{center}
\begin{tabular}{|c|c|}
\hline
\textbf{Type}& \textbf{AvgTime(ms)/Tput(tps)} \\
\hline
Application position computation & $9000\pm1000$/-- \\
Ranging computation & 0.5/-- \\
Chaincode read target & $20\pm20$/73 \\
Chaincode read device & $50\pm20$/28 \\
Chaincode update device & $2100\pm20$/0.5 \\
Chaincode add observation & $2070\pm20$/0.5 \\
Chaincode updat. evidence, reputation and trust & $2080\pm20$/0.5 \\
Chaincode comput. position  & $2110\pm20$/0.5 \\
\hline
\end{tabular}
\label{tab3}
\end{center}
\end{table}

\subsection{Discussion}
\subsubsection{Correctness}
In terms of correctness and functionalities, the tests and simulations revealed the following positive aspects:
\begin{itemize}
\item The successful execution of all application and chaincode operations demonstrates the robustness and reliability of the system.
\item The synchronization achieved by the Gateway application, which waits for a certain number of observations from every device before proceeding with the computation, ensures consistency and accuracy in the processing of device observations.
\item The use of Ultra-Wide Band (UWB) radio technology in the CLOVES testbed ensures that the prototype devices network operates with precision and speed, facilitating accurate and rapid measurements.
\end{itemize}
On the downside, the tests and simulation revealed a few drawbacks:
\begin{itemize}
\item The simulation does not fully replicate the connection between the testbed network and the application, limiting the testing of certain system aspects.
\item Due to limited resources for testing, the applications and blockchain can be simulated and tested together in the prototype, while the device network, including the ranging process and encryption, is simulated separately. This limitation restricts the testing of the IoT part of the system.
\item The network exhibits some rigidity post-initialization; while new devices can be created, there's no provision for device deletion, and only one target can be accommodated without the ability to add more.
\item Given the prototype nature of the tests, certain conditions of a real-world scenario are not accounted for.
\item The multilateration algorithm used assumes that the devices and target are all located in the same plane, which can limit its applicability in certain scenarios.
\end{itemize}
\subsubsection{Security and Privacy}
In terms of security and privacy, the tests and simulations revealed the following positive aspects:
\begin{itemize}
\item The successful execution of all application and chaincode operations demonstrates the security and privacy of the system.
\item This system mitigates the risk of malicious devices through trust calculation. However, its effectiveness relies heavily on the presence of numerous honest devices within the network.
\item This system mitigates the risk of malicious internal nodes through the PBFT consensus protocol. However, its effectiveness relies heavily on the presence of numerous honest nodes within the network.
\item This system ensures privacy by segregating data between different experiments using the collection concept. As a result, a client or peer cannot access the data of devices and targets from an experiment to which they do not belong.
\item This system provides varying levels of privacy, distinguishing between users and admins through chaincode checks. Users have access only to target data, while admins have access to both target and device data, thanks to the collection concept.
\item The blockchain ensures the availability and integrity of the data, providing a secure and tamper-proof environment.
\item The data transmitted between devices and the gateway are encrypted and hashed, ensuring both integrity and confidentiality.
\end{itemize}
On the downside, the tests and simulation revealed a few drawbacks:
\begin{itemize}
\item The system currently lacks security measures during the ranging process between devices and the target.
\item The current implementation lacks an efficient and highly secure method to ensure the reliability of data transmitted between devices and the gateway. Currently, only the device and target ID of the arrived message are checked, which may not be sufficient for ensuring complete reliability.
\item The service's availability is impacted by its dependency on receiving data from all necessary devices to complete the target position calculation. A distributed denial-of-service (DDoS) attack targeting a single device can disrupt data transmission, effectively blocking the service. While this approach ensures the service's security and precision by considering all devices, it also introduces a vulnerability wherein the service can be stalled if any device fails to transmit data.
\end{itemize}
\subsubsection{Performance}
In terms of performance, the tests and simulations revealed the following:
\begin{itemize}
\item Good performance is achieved by utilizing ultra-wideband technology, which enables rapid ranging calculations.
\item The application's position computation interval is quite large, which can lead to significant desynchronization between the real and calculated positions of the target. To mitigate this, the application buffers the data received from devices, reducing the impact of differences in data arrival and target position calculation speeds. Additionally, the application attempts to parallelize operations such as adding observations and confidence, and updating evidence, trust, and reputation, which seems to improve the time for position computation. In fact, considering the sum of average times for all chaincode operations needed to calculate the target, it amounts to approximately 14 seconds, which is less than the average time calculated from the application, 9 seconds.
\end{itemize}

\section[Related Work]{Related Work}
\label{related-work}

In recent years, numerous papers have been published discussing the integration of IoT and blockchain technologies. Additionally, various studies have focused on privacy within blockchain systems, with many of these addressing solutions involving Hyperledger Fabric.

``Trustworthy IoT: An Evidence Collection Approach based on Smart Contracts'' by Ardagna et al. \cite{b4} presents a service-oriented methodology leveraging blockchain and smart contracts for reliable evidence collection in IoT, balancing trustworthiness and performance, and validated with Hyperledger Fabric.
``PrivChain: Provenance and Privacy Preservation in Blockchain-enabled Supply Chains'' by Malik et al. \cite{b5} introduces a blockchain solution using Zero-Knowledge Succinct Non-Interactive Argument of Knowledge (ZK-SNARKs) to share proofs instead of raw data, with smart contracts automating proof verification and a monetary incentive mechanism for participants.
``Research on Distributed Blockchain-Based Privacy-Preserving and Data Security Framework in IoT'' by Tian et al. \cite{b6} presents a framework leveraging gateway devices, a middleware server, and blockchain integration to create a secure, privacy-focused P2P network for IoT data, utilizing Hyperledger Fabric for enhanced data privacy and encryption.
``AI-enabled Blockchain: An Outlier-aware Consensus Protocol for Blockchain-based IoT Networks'' by Salimitari et al. \cite{b7} introduces an AI-enabled blockchain (AIBC) network with a two-step consensus protocol combining outlier detection and Practical Byzantine Fault Tolerance (PBFT), integrating edge computing, AI, and blockchain for secure and efficient data processing, featuring a novel Honesty-Based Distributed Proof of Work (DPOW) mechanism to enhance security in public blockchain-IoT applications.
``A blockchain-based trust management method for Internet of Things'' by Wu et al. \cite{b8} introduces a blockchain-based trust management mechanism (BBTM) that uses mobile edge nodes to evaluate sensor node trustworthiness and facilitates smart contracts for trust computation and verification.

Many of these papers introduce the concept of computing the trustworthiness of devices on edge nodes. For example, in our project, we employ the same strategy as outlined in \cite{b8} using Hyperledger Fabric. It is notable that \cite{b4}, \cite{b5}, \cite{b6}, and \cite{b7} also utilize Hyperledger Fabric. In most cases, Hyperledger Fabric is used for its permissioned blockchain capabilities, however the works do not leverage its private data features. In \cite{b6}, different channels within the Fabric blockchain are used to privatize data. In our work, we present and utilize the concept of collections in Hyperledger Fabric. This approach enhances flexibility and reduces memory and cost requirements. Unlike deploying a new channel, which creates a new ledger, collections allow for logical separation of data within the same ledger. This makes the architecture more efficient and cost-effective.

\section{Conclusion}
\label{conclusion}

This paper introduces a system for localization of IoT devices in a privacy-preserving and trustworthy manner based on a permissioned blockchain. It builds upon the concepts outlined in the paper by Dedeoglu et al., which provides the foundation and principal method for managing the issue of malicious devices.
However, our system improves upon the work of Dedeoglu et al. in terms of privacy-preservation and access control.

The resulting prototype offers a solid foundation that can be readily downloaded, tested, and utilized to develop more realistic use cases. Despite its strengths, we observe that the prototype does have some limitations. These include the time required for target position calculation, the restriction to a single target, hypothetical simulation of communication between the gateway and devices, inflexibility in the structure, inability to remove devices from the blockchain, and the potential for the gateway main process to become non-reactive if a device stops transmitting. These limitations provide good directions for future work.

However, the project successfully achieves its primary goal of enhancing the privacy aspects through the concept of collections in Hyperledger Fabric. It introduces two levels of privacy: between localization experiments and between administrators and users. Additionally, the prototype implements device trust calculation to mitigate the risk of malicious devices, along with other techniques to enhance the system's trustworthiness, such as encrypting data transmitted between devices and the gateway.
Furthermore, the prototype implements verifiable processes for calculating the target's position, allowing administrators to manage the system, and enabling users to access target data.

\section*{Data Availability Statement}
The prototype code and data and be found on the project's Github page: \url{https://github.com/GuglielmoZocca/TrustLocalizationProject}.

\end{document}